\documentclass[12pt]{article}
\usepackage{times}
\usepackage{geometry}
\usepackage[utf8]{inputenc}
\usepackage{enumitem,amssymb}
\usepackage{ragged2e}
\usepackage{graphicx}
\usepackage{wrapfig}
\usepackage{subcaption} 

\newlist{thematic}{itemize}{8}
\setlist[thematic]{label=$\square$}
\usepackage{pifont}

\begin{document}
\raggedright
\huge
Astro2020 Science White Paper \linebreak

The Panchromatic Circumgalactic Medium \linebreak
\normalsize

\noindent \textbf{Thematic Areas:} \hspace*{60pt} $\square$ Planetary Systems \hspace*{10pt} $\square$ Star and Planet Formation \hspace*{20pt}\linebreak
$\square$ Formation and Evolution of Compact Objects \hspace*{31pt} $\square$ Cosmology and Fundamental Physics \linebreak
  $\square$  Stars and Stellar Evolution \hspace*{1pt} $\square$ Resolved Stellar Populations and their Environments \hspace*{40pt} \linebreak
  $\checkmark$    Galaxy Evolution   \hspace*{45pt} $\square$             Multi-Messenger Astronomy and Astrophysics \hspace*{65pt} \linebreak
  
\textbf{Principal Author: Q.\ Daniel Wang}

Institution: University of Massachusetts Amherst 
 \linebreak
Email: wqd@umass.edu
 \linebreak
Phone: 413-545-2131 
 \linebreak
 
\textbf{Co-authors:} 
  \linebreak
Joseph N. Burchett, Univ. of California - Santa Cruz\\
Nicolas Lehner, University of Notre Dame \\
John M. O'Meara, W. M. Keck Observatory\\
Molly S.\ Peeples, STScI / JHU\\
J.\ E.\ G.\ Peek, STScI / JHU\\
Marc Rafelski, STScI / JHU \\
Jason Tumlinson, STScI / JHU \\
Jessica Werk, University of Washington\\
Dennis Zaritsky, University of Arizona
\linebreak

\justifying

\textbf{Abstract  (optional):}
Galaxies are surrounded by extended atmospheres, which are often called the circumgalactic medium (CGM) and are the least understood part of galactic ecosystems. The CGM serves as a reservoir of both diffuse,
metal-poor gas accreted from the intergalactic medium, and metal-rich gas that is either ejected from galaxies by energetic feedback or stripped from infalling satellites. As such, the CGM is empirically multi-phased and complex in dynamics. Significant progress has been made in the past decade or so in observing the cosmic-ray/B-field, as well as various phases of the CGM. But basic questions remain to be answered. First, what are the energy, mass, and metal contents of the CGM? More specifically, how are they spatially distributed and partitioned in the different components? Moreover, how are they linked to properties of host galaxies and their global clustering and intergalactic medium environments? Lastly, what are the origin, state, and life-cycle of the CGM? This question explores the dynamics of the CGM.  Here we illustrate how these questions may be addressed with multi-wavelength observations of the CGM.

\thispagestyle{empty}

\pagebreak
\setcounter{page}{1} 
\justifying

\begin{figure}[!bth!]
\vskip -0.3cm
\begin{subfigure}[b]{0.48\textwidth}
    \includegraphics[width=\textwidth]{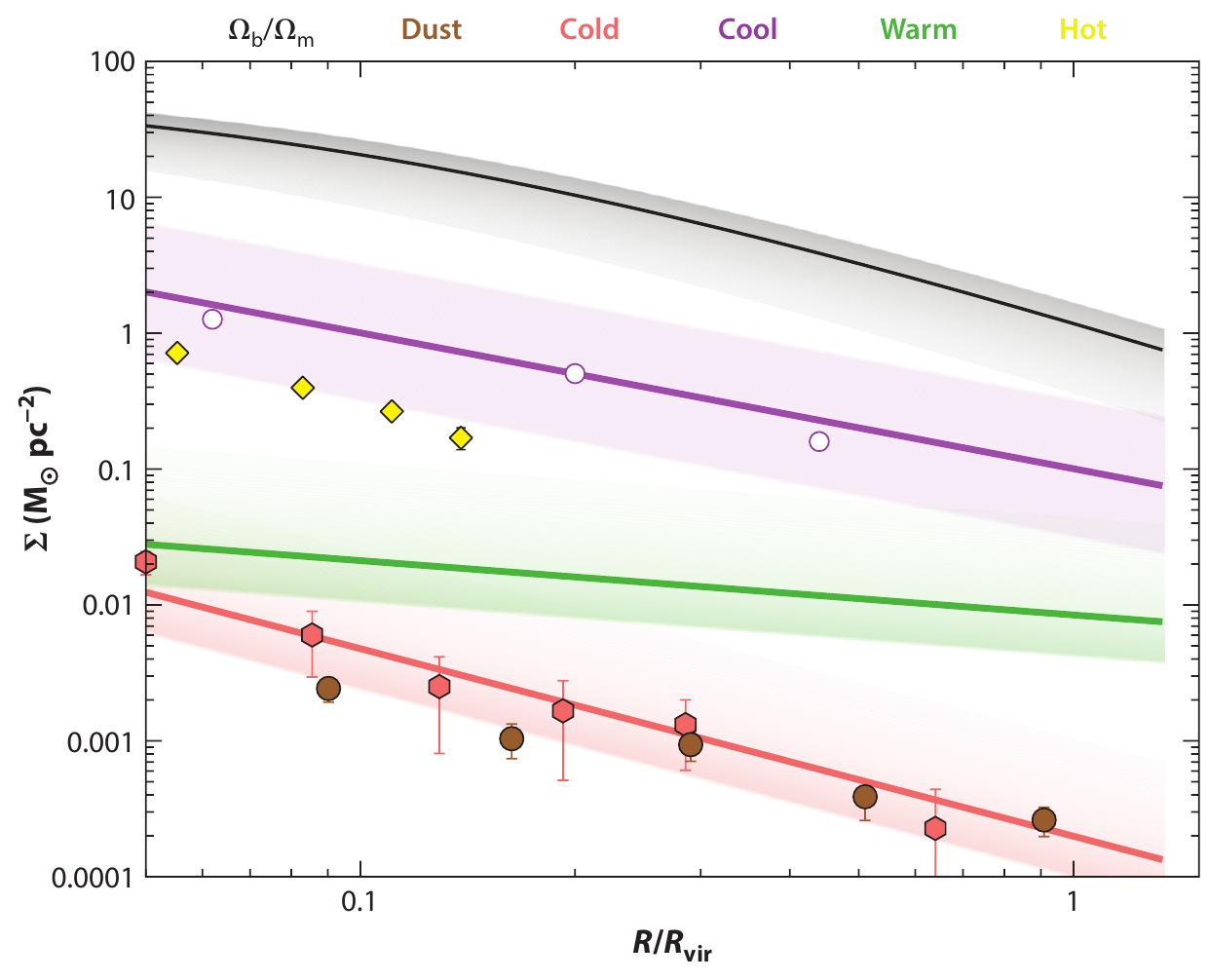}
  \end{subfigure}
  \begin{subfigure}[b]{0.5\textwidth}
    \includegraphics[width=\textwidth]{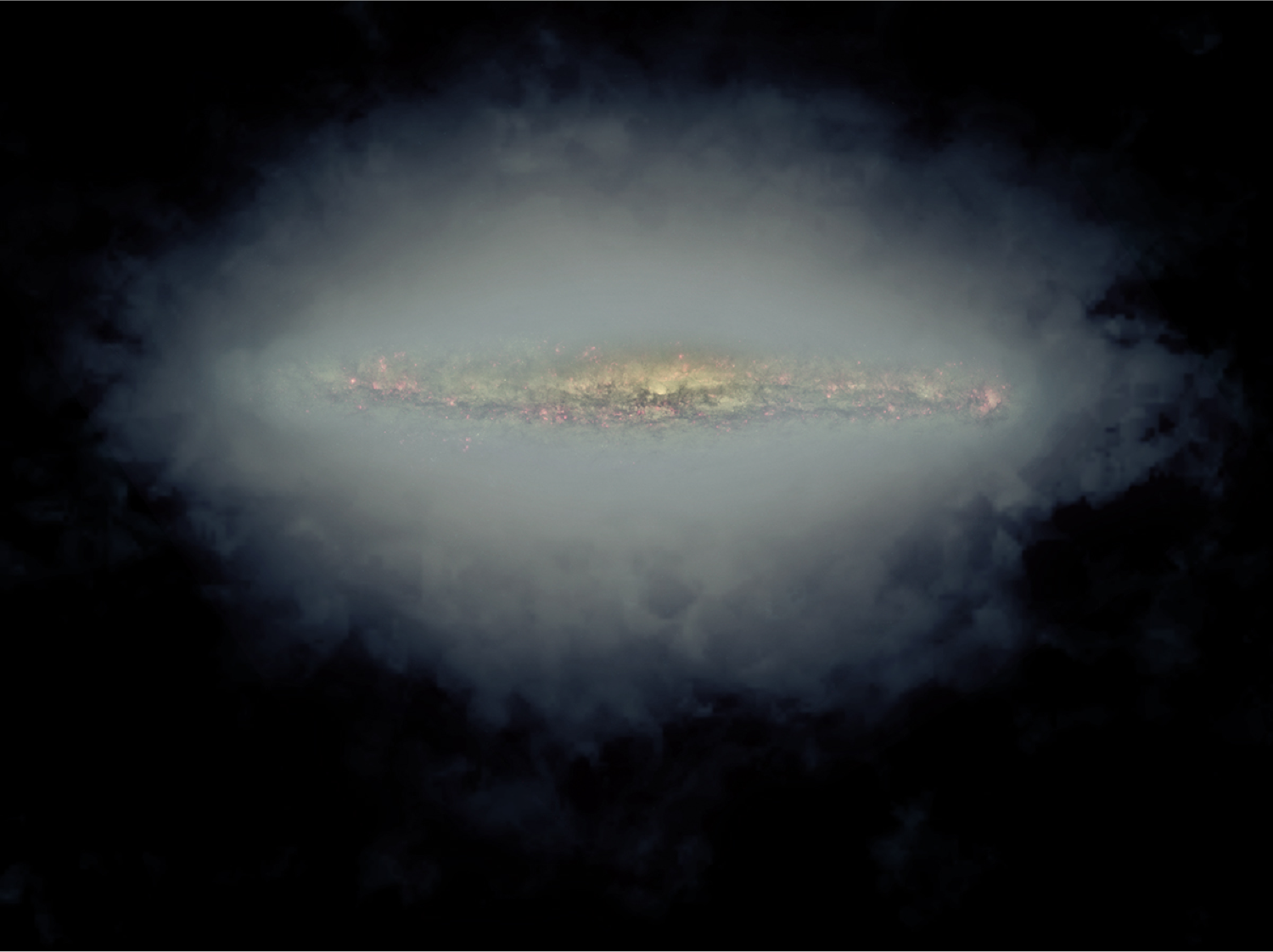}
  \end{subfigure}
  \caption{Left Panel: A multi-wavelength CGM mass density profile as of 2017 (Tumlinson, Peeples, \& Werk 2017). Right Panel: Median-averaged JVLA D-array/L-band radio halo of CHANGES galaxies (Wiegert et al. 2015).}
  \label{f:summary}
\vskip -0.3cm
\end{figure}

\begin{figure}[!th!]
\vskip -0.3cm
\centering
    \includegraphics[width=16cm]{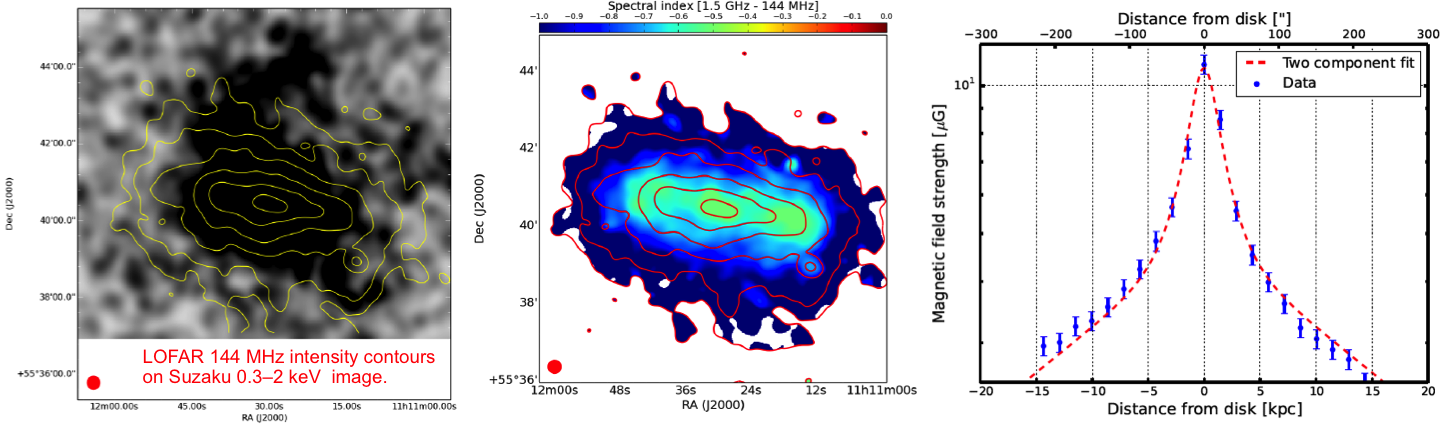}
    \caption{Highly inclinded disk galaxy NGC~3556 seen in radio and X-ray.}
    \label{f:n3556panel}
\vskip -0.3cm
\end{figure}

\noindent 
Galaxies breathe in and out of their extended atmospheres -- the circumgalactic medium (CGM). It occupies a vast volume from the disk of a galaxy to its virial radius and accounts for a substantial (if not dominant) fraction of its mass and energy, stored in various components (Table 1; Figs.~\ref{f:summary} and \ref{f:n3556panel}). Intensive multiwavelength efforts over the last decade have advanced our basic characterization of CGM gas from $\sim 1000$~K to $> 1$ million K, and in dust, such that only recently has it been possible to plot approximate mass density profiles for each component for galaxies near $L^*$ (Fig.~\ref{f:summary} left panel; 
Tumlinson, Peeples, \& Werk 2017). Observations in the X-ray, UV, and optical have all contributed to this diagram, and together show that the extended gaseous halos are a key part of the galaxy ecosystem. This accounting so far, however, remains greatly uncertain. Here we focus on how this research front has been advancing in recent years, pointing to key multi-wavelength observing capabilities that will help to expand this decade's progress to a wide range of galaxies over cosmic time, and to unravel not just the bulk properties and energy budgets but also the causal relationships between galaxies and their gas halos. 
We review progress and open questions specific for the hot component (\S\,1), magnetic fields and cosmic rays (\S\,2), the cool, warm, and dusty components (\S\,3), and conclude in \S\,4 with open questions primarily regarding the interplay among different components. Many of these questions can hopefully be addressed in the 2020s.

\begin{table}[!hbt!]\label{t:obs}
\begin{center}
\caption{CGM Components and Observing Methods}
\begin{tabular}{lcr}
\hline\hline
Component & Temperature (K) & Observing Method\\ 
\hline
hot gas & $10^{6}-10^{7}$ & soft X-ray emission/absorption, SZE (radio-submm)\\
warm gas & $10^{5}-10^{6}$&  Optical/UV line absorption/emission\\
cool gas & $10^{4}-10^{5}$&  Optical/UV line absorption/emission\\
cold gas &$< 10^{4}$& 21 cm line, Optical/UV line absorption\\
dust  & - & scattered UV, optical/UV extinction, far-IR to mm emission\\
CRs/B-field &- & radio continuum, hard X-ray, $\gamma$-ray\\
\hline		
\end{tabular} 
\end{center}
{\scriptsize 
CRs - cosmic rays;
SZE - Sunyaev?Zeldovich effect; IC - inverse Compton.
The ionized gas (cool-hot) and the B-field can also be studied statistically with fast radio bursts 
and extragalactic pulsars via observations of dispersion and Faraday rotation measures.
}
\end{table}

\noindent{\bf 1. Hot Component}
\medskip

Two principle methods can be used to probe the hot component of the CGM (Table 1). The SZE directly measures the overall thermal energy content of the CGM, which is likely dominated by the hot component.  However, SZE measurements for individual galaxies will typically be difficult, because of both the weakness of the signal and the cosmic variance of the CMB itself. Therefore, the SZE method will be most useful for statistical studies of the hot CGM. In contrast, soft X-ray observations of the hot CGM (Fig.~\ref{f:n3556panel} left panel; Li \& Wang 2013, Wang et al.\ 2016; Li et al.\ 2018) are mainly sensitive to its metal content. Soft X-ray spectroscopy, in particular, will become a powerful tool for probing various physical processes related to the life cycle of the hot CGM and its interface with cooler components. Over its expected temperature and metallicity ranges, the hot CGM can be studied via X-ray spectroscopy of emission/absorption lines from K-shell transitions of H-like and He-like ions, as well as those Fe L-shell ones (e.g., Wang 2010; Porquet et al.\ 2010; Foster et al.\ 2012). Studies typically assume an optically thin collisional ionization equilibrium (CIE) state for the hot CGM. However, because of its low density and hence long equilibration time scales, non-CIE processes should be important. They include the resonance scattering of various bright emission lines  (sensitive to the turbulent velocity of the hot CGM; e.g., Gu et al.\ 2016; Li \& Bregman 2017; Chen et al.\ 2018), charge exchange (CX) between ions and neutral atoms at hot/cool gas interfaces (e.g., Zhang et al.\ 2014), and over-heating (e.g., due to recent shock-heating) or over-cooling (such as fast adiabatic expansion of outflows in galactic disk/halo interfaces; e.g., Breitschwerdt \& Schmutzler 1999), and the photo-ionization by past or ongoing AGN activity (e.g., Segers et al.\ 2017). These processes must be understood to reliably infer the thermal and chemical properties of the hot CGM and to uncover relic information about recent AGN activity of host galaxies, for example. Existing studies use data from X-ray CCD and non-slit reflection grating spectrographs, all with very limited spectral resolutions.

High-resolution X-ray Imaging spectroscopy will become possible with the upcoming XRISM and Athena missions, although they are not optimized for soft X-ray observations. A dedicated soft X-ray spectroscopy mission, with combined capability for both emission (with a spectral resolution better than a couple of eV) and absorption (sub-eV), would truly make the astrophysical study of the hot CGM a reality.  We will then be able to map out individual emission lines in such key diagnostic line complexes as K$\alpha$ triplets of He-like ions, essential to the study of the non-CIE processes. Detailed studies of these processes, afforded at least for nearby galaxies (including the Milky Way), will help us to understand how the CGM works. For example, the soft X-ray line emission from the CX directly measures the relevant ion flux into hot/cold gas interfaces and can thus potentially provide a unique tool to measure their effective area and to trace stripped gas from satellite galaxies and/or cold inflows from the IGM. We can then address such questions as: Does the hot gas preheated in the IGM or is mostly ejected? And how does the hot gas lose its energy? 

For starburst galaxies, it will become possible to directly measure line-of-sight bulk and/or turbulent velocities for hot plasma in superwinds. So far, such velocity information is only available about the entrained cool gas, which shows bulk motion of velocity in the range of 200--1000\,km\,s$^{-1}$, as measured in UV/optical emission and/or absorption lines (e.g., Martin 2005). This velocity is expected to be substantially lower than that of  hot, metal-rich plasma in superwinds; but the relation between the two velocities is very uncertain (e.g. Strickland \& Stevens 2000). Therefore, to truly understand the mass rate, dynamics, and ultimate fate of superwinds, we must directly measure the velocity of hot plasma flows. The velocity can lead to the measurable centroid shift and/or width increase of the X-ray emission lines of such a flow. 

Equally important will be the X-ray absorption line spectroscopic capability. Unlike thermal X-ray emission, which is sensitive to the volume density square, absorption lines produced by ions directly probe their column densities, which are proportional to the mass of the hot plasma. The relative strengths of the absorption lines give direct diagnostics of the thermal, chemical and/or kinematic properties of the plasma. One can infer the velocity dispersion of the plasma, for example, from relative line saturation, even without resolving individual line profiles (e.g., Wang et al.\ 2005). Additional information can be extracted with the combination of the absorption line observation with the emission spectrum from the surrounding field. 

\medskip
\noindent{\bf 2. Cosmic-Ray/B-field Components}
\medskip

It is becoming increasingly clear that CRs, strongly coupled with B-field, play an essential role in regulating CGM structure and dynamics (e.g., Chan et al.\ 2018; Wiegert et al. 2015; Irwin et al.\ 2019). CRs are charged particles moving with relativistic
speeds, mainly generated through shock acceleration.  On average, about 10\% of the energy in a supernova remnant is expected to be carried away by CRs, mostly by protons. In the solar neighborhood the CR energy density is comparable to thermal and magnetic energy densities and is not consumed in the Galactic disk, suggesting that CRs escape into the CGM without losing much energy, at least in normal star-forming galaxies.

We are only starting to learn about the distributions of CRs and B-field in the CGM, mostly in regions close to galactic disks. The distributions can be traced by the synchrotron emission of CR electrons. As demonstrated by recent Continuum HAlos in Nearby Galaxies ? an EVLA Survey (CHANG-ES; Irwin et al.\ 2019), extended 
radio halos are ubiquitously seen around nearby highly inclined starforming galaxies (Fig.~\ref{f:summary} right panel and Fig.~\ref{f:n3556panel} left panel). Assuming energy equipartition 
between CRs and B-field, one finds that exponential scales heights of the latter are typically 1 to 2 kpc; the observed extent of such halos can be much greater(e.g., 8~kpc; e.g., Fig.~\ref{f:n3556panel} middle-right panels; Miskolczi et al.\ 2019; Krause et al.\ 2018). The CR transport speed estimated from the spectral 
steepening of the radio emission as a sign of the synchrotron aging suggest that the transport is 
mostly escape/advection-dominated even in many normal galaxies. With the typical advection speed of 
$(1-2) \times$ the escape velocity of the galaxies, CRs are surely escaping into the vast CGM space and possibly beyond. Interestingly,  the 
inferred B-field scale height (h) is correlated with the radio diameter (r) and the normalized scale height (h/r) 
decreases with the surface stellar mass density of a galaxy, indicating that its gravity can significantly slow down the transport. 
Furthermore, observations of radio polarization, assisted by the rotation measure synthesis 
technique, allows  for mapping the projected magnetic field directions. Several radio halos
have systematic reversals in the magnetic field direction (positive/negative) on kpc scales, providing 
powerful tests of dynamo theories (Irwin et al.\ 2019).

The relationship between the hot and CR/B-field components, as may be traced by soft X-ray and
radio continuum emission, has hardly been explored. While their overall luminosities are strongly correlated, there is little detailed {\em morphological} correlation between these two components. Ongoing detailed comparisons of high-resolution data indicates that radio ``chimneys'' tend to be filled with X-ray ``smoke'' in galactic disk/halo regions. In general, however, hot gas does not seem to be confined by a magnetic field, which may suggest that thermal pressure in hot gas is the main driving force of the outflows.

Clearly, more sensitive data are needed for a comprehensive exploration of the magnetized CGM, especially in its outer regions. Broad-band low frequency radio observing facilities allowing for decent spacing coverage will be most desirable to map out the radio emission to extremely low surface brightness, as well as to enable Faraday's rotation measurements against a statistically meaningful sample of background AGN for individual galactic radio halos. Of course, if IC-generated hard X-ray or soft $\gamma$-ray radiation from galactic radio halos can be measured, one can then infer both CR energy density and B-field intensity in the CGM without the equipartition assumption. Ultimately, we would 
like to directly probe high-energy CR protons and their interaction with the CGM via observations of $\gamma$-ray emission. The energy density of CRs peaks at GeV; their collisions with nuclei in the CGM produce pions, which decay into GeV-rays. Therefore, sensitive $\gamma$-ray observations will allow for constraints on the CR distribution and propagation (e.g., Chan et al.\ 2018). With these measurements, we will be able to address such questions as: How important are CRs/B-field to the overall energy budget? Could they even dominate the CGM? Can CR energy be transferred effectively to gas? Or can CRs be re-accelerated in the CGM? How significant are CRs as a source of non-thermal pressure, and could they be responsible for  the observed cool gas in the inner CGM of passive galaxies not accreting onto the ISM?


\medskip
\noindent{\bf 3. Cold, cool, and warm Gas and Dust Components}
\medskip

The thermal and ionization state of cool and warm gas (Table 1) in the CGM is still largely unconstrained (e.g., Tumlinson, Peeples, \& Werk 2017 and Zhang et al.\ 2018b for global diagnostics), but complex kinematic structures in absorption-line spectra indicate that physics beyond simple two-phase equilibrium models must be considered (Werk et al.\ 2014, 2017). Nevertheless, it is reasonably clear that gas at $\approx 10^{4}$--$10^5$K extends out to hundreds of kpc and is a dynamically and thermally important constituent of the CGM. Its contribution to the baryon budget of a 10$^{12}$ M$_{\odot}$ halo is $> 10^{10}$ M$_{\odot}$ and up to 10$^{11}$ M$_{\odot}$ (Werk et al.\ 2014, Stern et al.\ 2016, Keeney et al.\ 2017, Prochaska et al.\ 2017, Zhang et al.\ 2018a). Complementing the extensive work on absorption lines (see the review by Tumlinson, Peeples, \& Werk 2017), there is recent progress in the detection and analysis of emission lines from the cool gas (Zhang et al.\ 2016, 2018a, 2018b). These observational prospects are described in other Astro2020 white papers (Zaritsky et al., Tuttle et al.) but depend on the development of large ground based optical telescopes and UV observing capabilities.

Dust is also an important ingredient in the astrophysics of the CGM. Dust is observed out to 150\,kpc in stacked extinction maps using (Zaritsky 1994; Menard et al.\ 2010) and makes up half of the dust in the present day universe. It is also detected via diffuse ultraviolet light around individual highly inclined late-type galaxies out to 5--20 kpc from their midplanes (e.g., Hodges-Kluck \& Bregman 2014). The detected ultraviolet light traces the radiation escaping from the underlying galaxies and scattering off dust in the galactic halos. While dust provides a much smaller contribution to a galaxy's total baryon budget, it is significant in terms of the metals it contains relative to the metals produced by a galaxy over its lifetime (Peeples et al.\ 2014). In the relatively low temperature and low density CGM, dust can survive for long time (Peek et al.\ 2015) and can be potentially a major heat sink of hot gas. Dust provides a powerful complement to absorption line studies---it is not sensitive complex ionization physics, but is sensitive instead to the violence of ejection from galactic disks. High precision, large area unbiased photometry and high precision spectroscopy will enable detailed CGM dust studies in the 2020s and therefore place novel constraints on how material is ejected from galaxies.

Several directly relevant questions still need to be addressed: How important is the the CGM for stripping gas from satellite galaxies, and how much of the CGM is composed of this stripped gas? Does it lead to condensation or accelerated cooling of surrounding hot gas (McQuinn and Werk 2018)? How far does the cool, neutral gas in the CGM of passive galaxies extend, and how is this gas prevented from accreting onto quenched galaxies? What do the detailed kinematics of the multiphase suite of absorbing ions tell us about the physical and thermodynamic structure of the CGM? Are large-scale shocks responsible for the thermal and ionization state of the gas? 

\medskip
\noindent{\bf 4. Open Questions for Studying the CGM as Part of Galaxy Ecosystems in the Next Decade} 
\medskip

Clearly, progress has been made over the last decade, demonstrating the observationally feasibility of exploring the properties of various components of the CGM. In the meantime, key relevant physical progresses have been studied theoretically (e.g., Wiener et al.\ 2019) and have been implemented in state-of-the-art simulations of galaxy formation (e.g., Chan et al.\ 2019). 
Yet, while the basic contours of the CGM have been outlined, the underlying causal relationships are yet to be discovered. Here are some open questions about the CGM in the context of galaxy evolution and requiring a panchromatic approach to be addressed in the coming decade: 

{\bf What are the overall energy, mass, and metal contents of the CGM as a function of mass and time?} The cool, warm, hot, dust, and non-gaseous parts of the CGM have all been characterized to some extent, but the characterization has been focused, mostly as a result of observational limitations, on mainstream galaxies near L*. To isolate the causal effects of the CGM on galaxy evolution, we need to extend this full characterization to a wide range of mass, galaxy type, and star formation properties, from dwarfs to massive ellipticals. We need this in all phases of evolution, and in a range of environments within the large-scale structure. We need to understand how the phases are distributed spatially, and how the partitioning across the temperature regimes depends on the mass and star formation rate of galaxies. We also need to constrain how important cosmic rays and magnetic fields might be to the energy partition, which are now completely unconstrained. 

{\bf What are the origin, state, and life-cycle of the CGM?} Is most of the gas in the CGM at any time from IGM accretion, satellite galaxies, or ejected outflows? How much of it recycles into the galaxy, and on what timescales? Are the metals produced by, but missing from, galaxies, stored in the CGM? Do they recycle? Is the CGM more or less in hydrostatic equilibrium so pressure balance may be assumed? How important are bulk motions such as outflows and winds on global scales? How do the ``phases'' evolve into one another? How are ionization states determined in various phases? How important is the photo-ionization by AGN, even after they are off (long recombination time scales; Oppenheimer \& Schaye 2013; McQuinn \& Werk 2018)? Can CR energy be transferred effectively to gas? Or can CRs be re-accelerated in the CGM?

\pagebreak   
  
\vspace{0.2in}
   
\textbf{References}

Breitschwerdt, D. \& Schmutzler, T. 1999, A\&A, 347, 650

Chan, T.K. et al. 2018, arXiv:1812.10496

Chen, Y. Wang Q.D. et al. 2018, ApJ, 681, 138

Foster, A. R., Ji, L., Smith, R. K., \& Brickhouse, N. S. 2012, ApJ, 756, 128 

Gu, L. et al. 2016, A\&A, 594, 78

Hodges-Kluck, E., \& Bregman, J.N. 2014, ApJ, 789, 131

Irwin, J. et al. 2019, in the special issue of Galactic magnetism edited by Fletcher and Mao

Keeney, B.A., et al. 2017, ApJS, 230, 6

Krause, M., et al. 2018, A\&A, 611, 72 

Li, J.-T. \& Wang Q.D. 2013, MNRAS, 428, 2085

Li, J.-T. et al. 2018, ApJ, 855, 24

Li, M., Bryan, G.L.,  Ostriker, J.P. 2017, ApJ, 841, 101

Li, Y. \& Bregman, J. 2017, ApJ, 849, 105

Martin, C.L. 2005, ApJ, 621, 227

Menard, B., Scranton, R., Fukugita, M., \& Gordon, R. 2010, MNRAS, 405, 1025

Miskolczi, A., et al. 2019, A\&A, 622, 9

Oppenheimer, Benjamin D. \& Schaye, Joop 2013, MNRAS, 434, 1063

Peek, J. E. G., Menard, B., Corrales, L., 2015, ApJ, 813, 7

Porquet, D., Dubau, J., \& Grosso, N. 2010, SSRv, 157, 103

Prochaska, J.X., et al. 2017, 837, 169

Segers, M.C. et al. 2017, MNRAS, 471, 1026

Stern, J., Hennawi, J.F., Prochaska, J.X., \& Werk, J.K. 2016, ApJ, 830, 87

Strickland, D.K. \& Stevens, I.R. 2000, MNRAS, 314, 511 

Tumlinson, J., Peeples, M. S., \& Werk, J. K. 2017, ARAA, 55, 389 

Wang, J. et al. 2010, ApJL, 719, 208

Wang, Q.D. et al. 2005, ApJ, 635, 386 

Wang, Q.D. 2010, PNAS, 107, 7168

Wang, Q.D. et al. 2016, MNRAS, 457, 1385

Werk, J., et al. 2014, ApJ, 792, 8

Werk, J., et al. 2016, ApJ, 833, 54

Wiegert, T., et al. 2015, AJ, 150, 81

Wiener, J., Zweibel, E. G., \& Ruszkowski, M. 2019,  arXiv:1903.01471

Zaritsky, D. 1994, ApJ, 435, 599

Zhang, H., Zaritsky, D. \& Behroozi, P. 2018a, ApJ, 861, 34

Zhang, H., Zaritsky, D., Werk, J., \& Behroozi, P. 2018b, ApJL, 866, 4

Zhang, S. Wang Q.D. et al. 2014, ApJ, 794., 61

\end{document}